\documentclass[3p]{elsarticle}

\begin{document}

\title{Semi-analytic method for slow light photonic crystal waveguide design}
\author{Patrick Blown, Caitlin Fisher, Felix J. Lawrence, Nadav Gutman and C. Martijn de Sterke}
\address{ARC Center for Ultrahigh-bandwidth Devices for Optical Systems (CUDOS) and Institute of Photonics and Optical Sciences (IPOS),
School of Physics, University of Sydney, New South Wales, 2006, Australia}

\begin{abstract}
We present a semi-analytic method to calculate the dispersion curves and the group velocity of photonic crystal waveguide modes in two-dimensional geometries. We model the waveguide as a homogenous strip, surrounded by photonic crystal acting as diffracting mirrors. Following conventional guided-wave optics, the properties of the photonic crystal waveguide may be calculated from the phase upon propagation over the strip and the phase upon reflection. The cases of interest require a theory including the specular order and one other diffracted reflected order. The computational advantages let us scan a large parameter space, allowing us to find novel types of solutions. 
Keywords: photonic crystal waveguide; slow light; tailored dispersion; dispersion engineering

\end{abstract}

\maketitle

\section{Introduction}

Slow light has a number of important applications \cite{slowlightbook,Baba,Krauss}, many of which stem from the strong electric field which is associated with a particular energy flow. A key geometry for the generation and observation of slow light are photonic crystal waveguides (PCWs) in which the dispersion relation, and hence the group velocity, can be tuned very precisely.

One of the challenges in this field is the design of a structure with desired dispersion characteristics, typically involving both the group velocity and the dispersion. These can be engineered by varying the parameters of the first rows of the photonic crystal (PC) closest to the waveguide, for example the holes' radii or their transverse or lateral positions \cite{Schulz}. In this way, waveguide modes with low group velocity which weakly depend on frequency have been found \cite{White,Frandsen}. The actual design procedure is often based on a combination of experience and fully numerical optimization. This optimization typically starts with a two-dimensional geometry, which is followed by full three-dimensional slab geometry simulations in the most promising part of parameter space. The properties of the three-dimensional geometry, such as bandwidth and group index, are well predicted by those of the two-dimensional geometry \cite{White}.

\begin{figure}[htb]
\centering
\includegraphics[height=60mm]{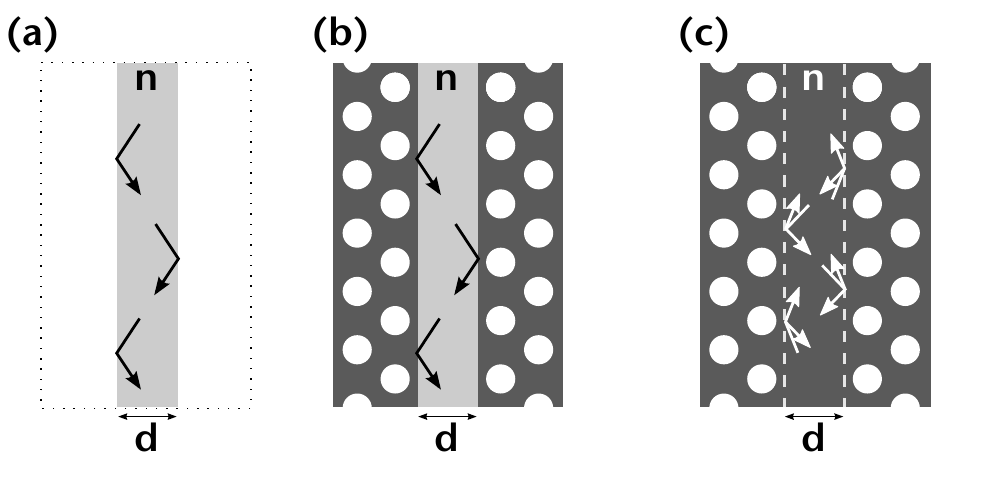}
\caption{Schematics of the two-dimensional geometries discussed. (a) Conventional planar geometry in which the mirror reflectivity stems from total internal reflection. (b) Scalar treatment of PCW in which the PC mirrors only reflect the specular order. (c) Vector treatment of PCW, in which the PC mirrors reflect the specular order and one diffracted order.}
\label{fig:schematic}
\end{figure}

Here we consider an efficient and conceptually convenient method for the calculation of the dispersion relation of waveguides in two-dimensional PCs. The key to our approach is that the PCW is considered to be a uniform strip which is surrounded by two mirrors, similar to a conventional planar waveguide as illustrated in Fig.~\ref{fig:schematic}. The key difference, of course, is that in planar waveguides the mirrors rely on total internal reflection, whereas in a PCW they rely on a combination of Bragg reflection from the PC mirrors and total internal reflection from the PCW's faces. Nonetheless, as we show here, the type of methods developed to analyze planar waveguides can be brought to bear on the calculation of the properties of PCWs. This leads to a numerical method which is efficient in that the width of the waveguide may be varied with little computational cost, and which provides some physical insight.
The design method for planar waveguides has been applied to PCWs previously. 
Sauvan and Lalanne used it to explain the PCW loss due to a finite cladding \cite{Sauvan}. 
Istrate and Sargent concentrated on PCW-based devices such as directional couplers and cavities \cite{Istrate}. 
Chen et. al. used it to determine the modes of a waveguide with a photonic crystal cladding and an air cover \cite{Chen}. 
None of these authors carried out a type of systematic slow-light analysis reported here. 

In Sec.~\ref{sec:Method} we formulate the theory. In Sec.~\ref{sec:scalar} we start with the case where only the specularly reflected order propagates, which applies to low-refractive index PCWs. As the PC mirrors serve as gratings, we formulate in Sec.~\ref{sec:vector} the theory when the PC mirrors reflect two propagating orders, which can be applied to practical PCWs. Some of the results for this are derived in the Appendix. In Sec.~\ref{sec:Results} we demonstrate the utility of our method and compare it to full three-dimensional numerical simulations. In Sec.~\ref{sec:Conclusions} we discuss and conclude our findings.

\section{Description of the method}\label{sec:Method}
\subsection{Low-refractive index waveguides--scalar case}
\label{sec:scalar}

It is well known that the resonance condition for a mode in a planar waveguide (see Fig.~\ref{fig:schematic}(a)) is given by \cite{Kogelnik}
\begin{equation}
   2nkd \cos\vartheta + 2\varphi = 2m\pi,
\label{FP}
\end{equation}
where $d$ is the width of the waveguide, $n$ is its refractive index, and $k=\omega/c$ is the wave number of the light in vacuum, with $\omega$ the frequency and $c$ the speed of light in vacuum. Further, $\vartheta$ is the angle of the wave with the normal of the mirror and $\varphi$ is the phase shift upon reflection ({\sl i.e.}, $r=\exp(i\varphi)$). Equation~(\ref{FP}) expresses the fact that the existence of a mode requires that the field is unchanged when traveling back and forth between the mirrors, which includes the phase accumulated due to the propagation and that due to two reflections off the mirrors, corresponding to the first and second terms on the left-hand-side of Eq.~(\ref{FP}), respectively (see also the Appendix). The standard treatment for planar structures then continues by substituting the expression for $\varphi=\varphi(\vartheta)$ for total reflection which follows from the Fresnel coefficients, to find a transcendental equation for the waveguide modes. In contrast, in our treatment for PCWs (Fig.~\ref{fig:schematic}(b)) we obtain $\varphi=\varphi(\vartheta,\omega)$ from a numerical calculation. We are also not only interested in the dispersion relation, but also in the group velocity. To this end we note that the propagation constant $\beta=nk\sin\vartheta$. The group velocity can thus be written as
\begin{equation}
   n_g\equiv{c\over v_g} = n \left(\sin\vartheta + \cos\vartheta \, \left(k{d\vartheta\over dk}\right)\right).
\label{vg}
\end{equation}
 We now implicitly differentiate Eq.~(\ref{FP}) with respect to $k$ to find the expression for $kd\vartheta/dk$
\begin{equation}
   k{d\vartheta\over dk}={nd\cos\vartheta +{\partial\varphi\over \partial k}|_\vartheta\over n d\sin\vartheta-{1\over k}{\partial\varphi\over \partial\vartheta}|_k}
   \equiv {d_t\cos\vartheta\over d_w \sin\vartheta}.
\label{scalar}
\end{equation}
The key elements, and the only elements which depend explicitly on the properties of the mirrors, are the partial derivatives of the phase upon reflection with respect to frequency and incident angle. The phase shift for reflection off an ideal metal is $\pi$, and thus $d_t=d_w=nd$ for a metal waveguide. In the most general situation the parameters $d_w$ and $d_\tau$ can be interpreted as follows: the partial derivative $\partial\varphi/\partial\vartheta$ enters the expression for the Goos-H\"anchen shift, a sideway shift if the incident beam has a finite width, and which can be thought of as a correction to the position of a mirror \cite{Kogelnik}. Thus, $d_w$ is the waveguide width corrected for the Goos-H\"anchen shift. The inclusion of the factor $n$ indicates that $d_w$ corresponds to an optical path length. Similarly, the partial derivative $\partial\varphi/\partial\omega$ is associated with the group delay, an apparent shift to the position of the mirror stemming from finite pulse lengths \cite{Knox}, and $d_\tau$ is the waveguide width corrected for this effect.

From Eqs~(\ref{vg}) and~(\ref{scalar}) we see that $v_g\rightarrow0$ when the denominator in~(\ref{scalar}) vanishes. Since the first term in the denominator in~(\ref{scalar}) is positive, this means that $\partial\varphi/\partial\vartheta$ needs to be small and positive. In regular dielectric waveguides $\partial \varphi/\partial\vartheta$ is negative and so the group velocity does not vanish. In PCWs, $v_g=0$ both at the edge and at the middle of the Brillouin zone (BZ). We have found the general conditions for which $v_g=0$ by considering the properties of the mirror only, but we do not consider these here, since, as discussed in the next paragraph, they apply only to waveguides made of low-index material.

Results (\ref{vg}) and~(\ref{scalar}) are interesting but we have implicitly assumed the wavenumber in the waveguide to be small. This arises from the assumption that an incoming plane wave leads to a single reflected plane wave as in Fig. \ref{fig:schematic}(b). However, each of the sides of the waveguide acts as a reflection grating, which, at sufficiently high frequencies, leads to diffracted orders, the angles of which are given by the grating equation \cite{Istrate}. For PC bandgap mirrors, result~(\ref{scalar}) is valid only for an air waveguide, or waveguide made of another low-index material, sandwiched between the two PC mirrors. For the key case in which the waveguide is made of silicon or another high-index material, the frequencies for which Eq.~(\ref{scalar}) holds correspond to a band, rather than to the PCs' photonic band gap. To treat the more interesting case it is sufficient to consider frequencies at which the mirror has one diffracted order in addition to the specularly reflected order, as discussed in Sec.~\ref{sec:vector}.

\subsection{High-refractive index waveguides--vector case}
\label{sec:vector}
Here we consider the case in which the mirrors have two reflected orders (labeled $0$ and $-1$), the specular order and one other (see Fig.~\ref{fig:schematic}(c)). In this situation the reflection coefficient is replaced by a $2\times2$ reflection matrix, in which the diagonal elements give the specular reflection of the two orders, and the off-diagonal elements their coupling. It would seem that in the most general case this matrix is characterized by four complex numbers, but by imposing reciprocity this number is limited to four real numbers, and the general form of the matrix is
\begin{equation}
R = \left(
  \begin{array}{cc}
    a e^{i\varphi_0} & \pm i {1-a^2\over b} e^{i(\varphi_0+\varphi_{-1})/2}  \\
    \pm ib e^{i(\varphi_0+\varphi_{-1})/2} & a e^{i\varphi_{-1}} \\
  \end{array}
\right),
\label{matrix}
\end{equation}
where $a$, $b$, and $\varphi_{0,-1}$ are real. Energy conservation provides a relationship between $a$ and $b$, but we do not need this here. 
Following the same approach as in the scalar case, we derive in Appendix A the transcendental equation for the odd and even fundamental modes. 
The approach is similar to that of the calculation of Lamb waves in acoustics \cite{Auld}, which also involve two plane waves that couple at the mirrors.  The resulting equation for triangular lattices reads
\begin{equation}
   \tan\psi_0\cot\psi_{-1} = {a \mp 1 \over a \pm 1},
\label{Lamb}
\end{equation}
where $\psi_m = (nkd\cos\vartheta_m+\varphi_m)/2$, and the upper and lower signs stand for the even and odd modes respectively. 
For square and rectangular lattices, $\cot \psi_{-1}$ is replaced by $\tan \psi_{-1}$.

The next step, which is tedious but straightforward, is to implicitly differentiate  Eq.~(\ref{Lamb}) to find 
\begin{equation}\label{eq:Matrix}
   k{d\vartheta_0\over dk}= {{1\over\sin 2\psi_0}\left(nd\cos\vartheta_0 +{\partial\varphi_0\over\partial k}\right)- {1\over\sin 2\psi_{-1}}\left(nd\sec\vartheta_{-1}-nd\tan\vartheta_{-1}\sin\vartheta_0 +{\partial\varphi_{-1}\over\partial k}\right)-{2\over a^2-1}{\partial a\over\partial k}
   \over {1\over\sin 2\psi_{0}}\left(nd\sin\vartheta_0 -{1\over k}{\partial\varphi_0\over\partial \vartheta_0}\right)-
   {1\over\sin 2\psi_{-1}}\left(nd\tan\vartheta_{-1}\cos\vartheta_0 +{1\over k}{\partial\varphi_{-1}\over\partial\vartheta_0}\right)+{2\over a^2-1}{1\over k}{\partial a\over \partial \vartheta_0}},
\end{equation}
where $\vartheta_0$ and $\vartheta_{-1}$ are related to each other by the grating equation. Note that parameter $b$ does not enter this equation. 
Finally, Eq.~(\ref{eq:Matrix}) may be used in Eq.~(\ref{vg}) to express the group velocity in terms of the elements of the reflection matrix~(\ref{matrix}). 
Even though there are two plane-wave orders, by the grating equation the associated $\beta = nk\sin\vartheta$ differ by a reciprocal lattice vector, and so they lead to the same group index when evaluating $k{d\vartheta\over dk}$.

The first terms of Eq.~(\ref{eq:Matrix}) in the numerator and the denominator are the same as in Eq.~(\ref{scalar}), the scalar case. The second terms are similar, but for the $-1$ order: as the incident angle of the zeroth order plane waves varies, the $-1$ order varies as well, since the angles are related by the grating equation. The third terms are new and lead to a phase distortion, rather than an amplitude distortion, of the orders, but we have found these to be small in practice. As in the previous section, where there is only one specular reflection, $v_g=0$ when the denominator in (\ref{eq:Matrix}) vanishes.

Our method can be used to show that $v_g = 0$ at the BZ edge; we now briefly discuss how. 
By spatial symmetry, the plane wave orders constituting the backwards waveguide mode (denoted by a prime) are related to those comprising the forward mode: $\vartheta^\prime_0 = - \vartheta_{0}$ and $\vartheta^\prime_{-1} = -\vartheta_{-1}$, where for consistency we have also reversed the grating's period, $\Lambda^\prime = -\Lambda$. 
At the BZ edge, the two grating orders are in Littrow configuration, i.e., $\beta_0 = \pi/\Lambda$, $\beta_{-1} = -\pi/\Lambda$, and $\vartheta_0 = - \vartheta_{-1}$. 
Therefore at the BZ edge, $\vartheta^\prime_0 = \vartheta_{-1}$ and $\vartheta^\prime_{-1} = \vartheta_0$: the backward and forward waveguide modes are composed of the same two plane waves. 
Applying symmetry properties and energy conservation to Eq.~(\ref{matrix}), we see that the off-diagonal terms of $R$ are equal, and that $\varphi_0 = \varphi_{-1}$ for rectangular lattices and $\varphi_0 = \varphi_{-1} + \pi$ for triangular lattices.
Also, $k_{0t} = k_{-1t}$, so by solving Eq.~(\ref{eq:half_round_trip}) we find that the amplitudes of the plane wave orders in a mode are related: $A_{-1} = \pm A_0$ for a square lattice, and $A_{-1} = \pm i A_0$ for a triangular lattice. 
We have therefore shown that two waveguide modes consist of the same two plane waves with the same ratio of amplitudes: thus they are indistinguishable. 
Since by definition the modes propagate in opposite directions, they must have $v_g = 0$.

\subsection{Calculation of $\varphi_0$ and $\varphi_{-1}$}
\label{sec:Mirror}
The waveguide mirrors that we consider are either semi-infinite PCs, or stacks of PC layers in front of semi-infinite PCs. The reflection matrix (\ref{matrix}) describes the amplitudes and phases of the two plane waves' reflections and diffractions off this mirror. The matrix, and thus $\varphi_0$ and $\varphi_{-1}$, may be calculated with electromagnetic scattering software by simulating the mirror with one incident plane wave at a time, and measuring the complex amplitudes of the reflected plane waves. We calculate the reflection matrix using $2 \times 2$ PC impedance matrices~\cite{Lawrence} found by an accurate and efficient multipole method \cite{Multipole}.

\section{Results}\label{sec:Results}
\begin{figure}[htb]
	\centering
	\includegraphics[width=12cm]{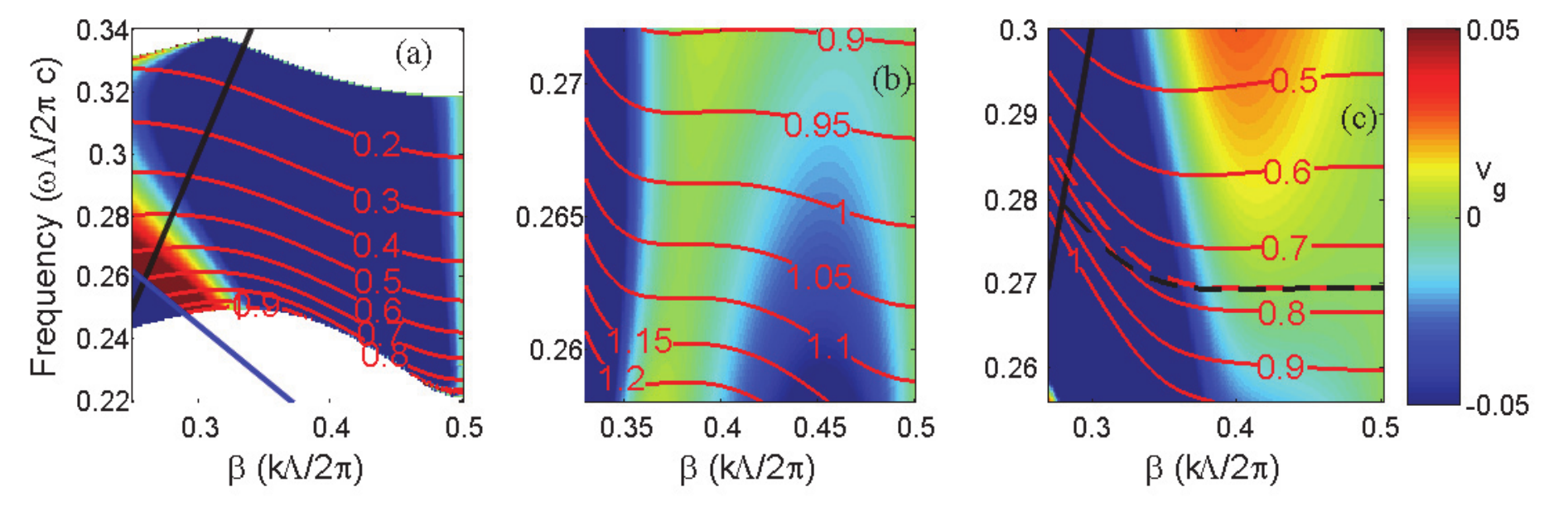}
	\caption{Dispersion relations for different PCW designs for varying waveguide thickness.
    The red curves give the waveguide width in units of $\Lambda\sqrt3/2$. (a) $H$ out of plane, $n=2.86$ and $r=0.3\Lambda$; black line: light line; blue line: Wood anomaly. (b) similar to (a) but with $r=0.404\Lambda$ in the second row of holes: the dispersion relation has a stationary inflection point for a W1 waveguide (width $d=\Lambda\sqrt3/2$). (c) Similar to (a), but $r=0.26\Lambda$ and $r=0.34\Lambda$ in the first and second rows of holes, respectively: the dispersion relation has a quartic degenerate band edge for $d=0.76\Lambda\sqrt3/2$.}
	\label{pic:results}
\end{figure}

We now apply the method from Sec.~\ref{sec:vector} to design a number of PCWs. The most important, and computationally slowest, part of our computation is determining the reflection characteristics of the mirrors. Since at the beginning of the calculation the frequency and incident angle are unknown, the parameters $a$, $\varphi_{0}$, and $\varphi_{-1}$ from Eq.~(\ref{matrix}) need to be calculated for all incident angles and all frequencies inside the bandgap. This also provides a key advantage: once these are known the dispersion relation for a waveguide of any width can be calculated with very little additional effort. To illustrate this consider Fig.~\ref{pic:results}(a), for which $r=0.3\Lambda$, where $r$ is the radius and $\Lambda$ is the period, $n=2.86$, and light is polarized with the $H$ field out of the PC plane. The curve labeled ``0.5'' in this figure, for example, shows the dispersion relation for the even mode of a W0.5 waveguide, {\sl i.e.}, a waveguide of width $0.5 \Lambda\sqrt3/2$. The other curves give the dispersion relations for waveguide of other widths. Using our method it is very easy to calculate the dispersion relations of waveguides of many different widths. This fills the entire bandgap, and the color coding gives the group velocity for the unique dispersion relation at each point. White regions correspond to bands, where the PC does not act as a mirror for that incident angle. The black line is the light line, whereas the blue line is the first Wood anomaly. Below the Wood anomaly the scalar treatment from Sec.~\ref{sec:scalar} is sufficient, whereas above the Wood anomaly the vector treatment (Sec.~\ref{sec:vector}) is required. The figure shows the expected behavior: the group velocity approaches zero as $\beta$ approaches the edge of the BZ.

In 2008, White {\sl et al.} \cite{White2008} published a waveguide design with a purely cubic dispersion relation, for a W1 waveguide. Near this stationary inflection point the waveguide exhibits two desirable properties: the modal group velocity is low and efficient coupling into this slow mode is possible \cite{White2008, Figotin2003}. The results for this design are shown in Fig.~\ref{pic:results}(b), which has the same parameters as in Fig.~\ref{pic:results}(a), except that for the second row of holes $r=0.404\Lambda$. 
Our results agree well with those of White \emph{et al.}: we obtain a similar dispersion curve for the W1 waveguide, including the presence of a stationary inflection point. 
Our figure also shows how the dispersion relation evolves with waveguide width.

The final example is a variation on the previous ones in that the radii of the first and second rows of holes were changed to $r=0.26\Lambda$ and $r=0.34\Lambda$ respectively. The resulting dispersion relations are summarized in Fig.~\ref{pic:results}(c). The figure shows that for narrow waveguides ($d<0.7\Lambda\sqrt3/2$) the dispersion relation curves down when moving away from the BZ edge, while for wider waveguides $d>0.8\Lambda\sqrt3/2$, it initially curves up. This implies that for waveguides of intermediate width the curvature close to the BZ edge vanishes. In other words, the dispersion relation is locally quartic. This is confirmed by the red dashed curve, which gives the dispersion relation for a waveguide of width $d=0.76\Lambda\sqrt3/2$. Like cubic dispersion relations, the evanescent modes associated with these features allow for efficient coupling into slow-light waveguides \cite{Gutman}. This example shows the advantages of being able to scan a part of parameter space with little computational effort.

The treatment here is a two-dimensional one in which the structure is taken to be uniform in the direction orthogonal to the periodicity. As mentioned, the design of PCWs often starts with a two-dimensional treatment to identify the most promising part of parameter space, which is then followed by a full three-dimensional treatment. The black curve in Fig.~\ref{pic:results}(c) shows the results of a fully-vectorial 3D eigenmodes solution of Maxwell's equations in a plane wave basis \cite{Johnson}, with a slab thickness of $0.58\Lambda$ and the refractive index of silicon ($n=3.5$). The presence of the quartic dispersion relation confirms that the two-dimensional calculations are an excellent predictor for full results.

\section{Discussion and Conclusions}\label{sec:Conclusions}
We have presented a semi-analytic method for calculating the dispersion relation of PCW modes, based on the analogy with conventional planar waveguides. This method is efficient in that calculating the dispersion for different waveguide widths requires negligible additional calculational effort. The most important limitation is that our treatment is a two-dimensional one, so that the results for three-dimensional geometries follow from an effective-index \cite{Kogelnik} argument. The reason for the two-dimensional treatment is that, even for conventional, non-circular waveguides ({\sl i.e.}, not based on PCs) the calculation of the modes for three-dimensional structures is usually fully numerical and is not cast in term of mirror properties. In fact the key exception to this is when the effective index method is used! The results in Fig.~\ref{pic:results}(c) confirm though that a two-dimensional treatment gives an excellent indication of the full, three-dimensional properties.

Other, more minor approximations are that we have assumed that the refractive index of the waveguide material does not depend on frequency. The medium dispersion is easily included but this small effect leads only to very small changes in the final results. We have also assumed the presence of two propagating diffracted orders: the specular order and one other. Though for the parameters of interest no additional propagating orders exist, a more complete theory would need to include evanescent orders. However, at the frequencies of operation the evanescent orders tend to decay significantly over the width of the waveguide, and, additionally, their amplitudes are in fact very small as well. Indeed, we have found that the vector theory from Sec.~\ref{sec:vector} is sufficient for our purposes. This is illustrated by the fact that our method reproduces the presence of the dispersion relation which locally is cubic (Fig.~\ref{pic:results}(b)), a delicate feature which is sensitive to the parameters and to the use of approximations.

\section*{Acknowledgements}
We thank Dr Tom White for useful discussions. This work was supported by the Australian Research Council.

\appendix
\section{Derivation of Eq.~(\ref{Lamb})}
In this Appendix we derive Eq.~(\ref{Lamb}) for a mode consisting of two coupled plane wave orders, using a resonance argument and expression~(\ref{matrix}) for the reflection off the interface with the photonic crystal. 
We do so following an argument set forth earlier for acoustic waves \cite{Auld} (see also \cite{Istrate}). 
Since the geometry is symmetric the solutions are either even or odd. 
Let us take the complex amplitudes of the two orders at the left of the waveguide, propagating to the right, to be $(A_{-1},A_0)$ for the $-1$ and $0$ orders, respectively. 
They then propagate to the right-hand mirror and undergo reflection as described by Eq.~(\ref{matrix}), before returning to the left-hand mirror and reflecting again. 
For a mode to exist, the phase accumulated in this process needs to be such that the amplitudes are $(A_{-1},A_0)$. 
We can rewrite this condition in vector form as
\begin{equation}
	\label{eq:full_round_trip}
   R \cdot P^- \cdot R \cdot P^+ \left(\begin{array}{c} A_{-1}\\ A_0 \\ \end{array}\right)= \left(\begin{array}{c} A_{-1}\\ A_0 \\ \end{array}\right),
\end{equation}
where $R$ is given by Eq.~(\ref{matrix}) and for square lattices
\begin{equation}
   P^+ = P^- =
   \left(
   \begin{array}{cc}
    e^{ik_{-1t}d} & 0  \\
    0 & e^{ik_{0t} d} \\
   \end{array}
   \right),
\end{equation}
where $k_{0,-1t}=k\cos(\vartheta_{0,-1})$ is the transverse component of the wavevector. 
Since $P^+ = P^-$ for square lattices, we factorize Eq.~(\ref{eq:full_round_trip}), writing $P = P^\pm$:
\begin{equation}
 	\label{eq:half_round_trip}
    R \cdot P \left(\begin{array}{c} A_{-1}\\ A_0 \\ \end{array}\right)=\pm \left(\begin{array}{c} A_{-1}\\ A_0 \\ \end{array}\right),
\end{equation}
consistent with the result for acoustic waveguides \cite{Auld}.

\begin{figure}[tb]
\centering
\includegraphics[height=39mm]{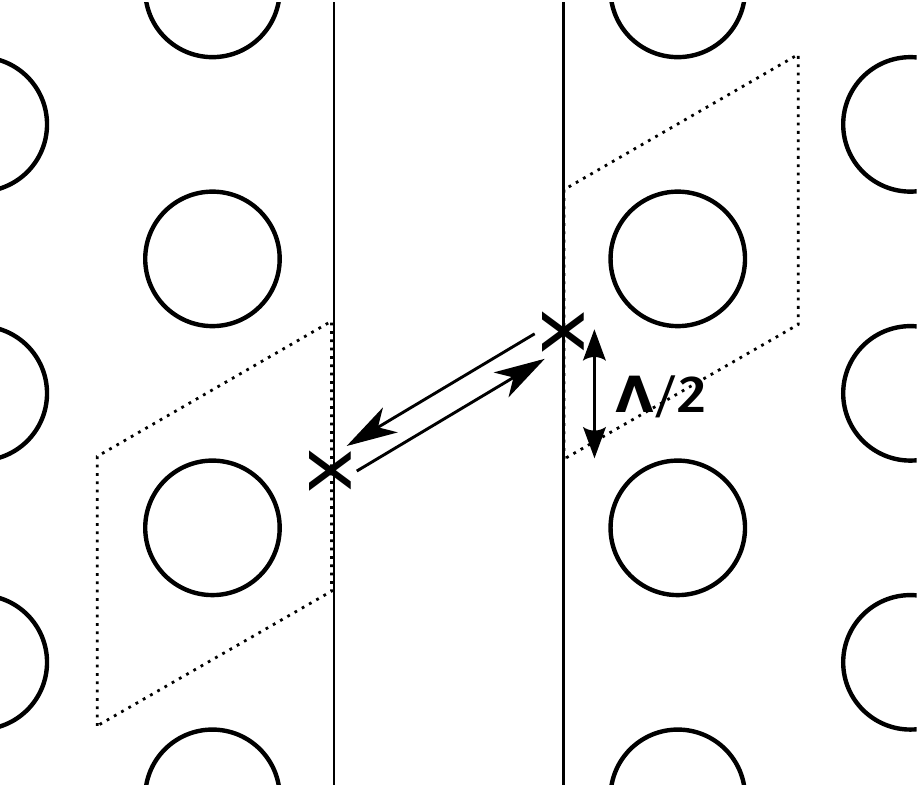}
\caption{The PC mirror's reflection matrix is calculated with respect to the point at the centre of the PC's unit cell's edge. This means that a phase shift associated with this vertical translation of $\pm \Lambda/2$ must be included in Eq.~(\ref{eq:half_round_trip}).}
\label{fig:ref-phase-origin}
\end{figure}

A subtlety applies to triangular lattices, which necessitates a modification to $P$.
The reflection matrix $R$ is defined with respect to a particular point---if this point is moved along the PC interface, then the phases of the off-diagonal elements change because $\beta_0 \neq \beta_{-1}$. 
For triangular lattices, which have non-orthogonal lattice vectors, the reflection matrices for the left and right mirrors are equal when they are defined for points vertically separated by a half-integer number of PC periods, e.g. $\Lambda/2$ (Fig.~\ref{fig:ref-phase-origin}). 
This vertical shift must be included in $P^\pm$ for Eq.~(\ref{eq:full_round_trip}) to apply to triangular lattices:
\begin{equation}
    P^\pm =
    \left(
    \begin{array}{cc}
     e^{i(k_{-1t}d \pm \beta_{-1}\Lambda/2)} & 0  \\
     0 & e^{i(k_{0t} d \pm \beta_0 \Lambda/2)} \\
    \end{array}
    \right).
\end{equation}
By the grating equation, $P^- = P^+ e^{-i\beta_0 \Lambda}$. 
Inserting this result into Eq.~(\ref{eq:full_round_trip}) and factorizing, we obtain Eq.~(\ref{eq:half_round_trip}) for triangular lattice PC mirrors but with
\begin{equation}
	P = 
	\left(
   \begin{array}{cc}
    e^{ik_{-1t}d} & 0  \\
    0 & -e^{ik_{0t} d} \\
   \end{array}
	\right).
\end{equation}
Some straightforward manipulation and an application of the grating equation then leads immediately to Eq.~(\ref{Lamb}), in which $b$ does not appear. In the presence of only the specular order the vectors and matrices become scalars and $a=1$, leading immediately to Eq.~(\ref{FP}).



\end{document}